\documentclass[preprint,12pt]{elsarticle}

\usepackage{graphicx}
\usepackage{amssymb}
\biboptions{compress}

\usepackage{ifthen}
\newboolean{ARXIV}
\setboolean{ARXIV}{true}
\newcommand{\arxiv}[1]{\ifthenelse{\boolean{ARXIV}}{ [{\tt #1}]}{}}
\newboolean{TITLES}
\setboolean{TITLES}{true}
\newcommand{\settitle}[1]{\ifthenelse{\boolean{TITLES}}{ \emph{#1}.}{,}}

\newcommand{\Tr}{\ensuremath{\mathrm{Tr}}}
\newcommand{\Real}{\ensuremath{\mathrm{Re}}}
\DeclareRobustCommand{\Eqref}[1]{Eq.~(\ref{#1})}
\DeclareRobustCommand{\Tabref}[1]{Tab.~(\ref{#1})}
\DeclareRobustCommand{\Figref}[1]{Fig.~(\ref{#1})}

\begin{document}

\begin{frontmatter}

\title{Finite temperature effective string corrections in $(3+1)D$ $SU(2)$ lattice gauge theory}
\author{Claudio Bonati}
\ead{bonati@df.unipi.it}
\address{Dipartimento di Fisica, Universit\`a di Pisa and INFN, Sezione di Pisa, \\ Largo Pontecorvo 3, 56127 Pisa, Italy}

\begin{abstract}
We study the effective string corrections to the inter-quark potential at finite temperature by simulating
the $SU(2)$ lattice gauge theory in four dimensions. We provide the first numerical evidence that the logarithmic 
correction to the potential, which was recently proposed to be a signature of the effective string at finite temperature, 
is universal also in $(3+1)D$ gauge theory, thus extending previous results limited to the $(2+1)D$ case.
\end{abstract}


\end{frontmatter}

\section{Introduction}
\label{into}

A precise understanding of the mechanism responsible for color confinement is still lacking, however there is by now 
little doubt that the interaction between a static quark-antiquark pair can be described by an effective string theory.  
From this point of view, initiated by the seminal studies \cite{LSW, L, LMW}, the presence of a flux tube joining the 
static quark-antiquark pair is given for granted and its physical consequences are examined.

At zero temperature the best known consequence of the effective string description is the presence of the so called 
``L\"{u}scher term'' \cite{LSW, L, LMW}, \emph{i.e.} the presence in the long distance inter-quark potential of a term 
proportional to $1/r$ ($r$ being the inter-quark distance), with a known coefficient which depends only on the dimension 
of the space-time (in particular it does not depend on the gauge group) and it is not influenced by higher order 
corrections of the effective string action. 
As a consequence of these general properties, the presence of the L\"{u}scher term in the inter-quark potential can be 
considered as a clear signature of the effective string behaviour of the flux tube. Because of its 
theoretical interest much studies have been devoted, in the field of lattice gauge theory simulation, to detect the 
presence of the L\"uscher term and to check its universal properties (see \emph{e.g.} \cite{LW02, CHP, CPR}).

The effective string picture can be used to describe the inter-quark potential also at nonzero temperature $T$ (that must 
clearly be lower than the deconfinement temperature $T_c$): in this regime the dominant string correction to the potential 
is linear in $r$ and decreases the value of the string tension. This is temperature dependent and ultimately vanishes 
at deconfinement.
In \cite{CFPP} it was recently argued that an universal signature of the effective string at finite temperature, 
analogous to the L\"{u}scher term at zero temperature, is given by the logarithmic correction to the inter-quark 
potential. This conclusion was supported by numerical evidence obtained by performing simulations in $D=2+1$ lattice 
gauge theories for various gauge groups. It was also noted that these numerical simulations are particularly time consuming, 
since the logarithmic correction is washed out by the projection to zero transverse momentum of the Polyakov loop, a common trick 
used in order to reduce the statistical errors (see \emph{e.g.} \cite{ABT}).
 
The aim of this paper is to show that the universal form of the logarithmic correction to the inter-quark potential at finite temperature
(as predicted by the effective string picture) is consistent with the numerically generated lattice gauge data also for the $D=3+1$ 
$SU(2)$ lattice gauge theory.

\section{The lattice setup and the effective string prediction}
\label{latticestring}

We studied the $SU(2)$ Yang-Mills theory discretized on an isotropic cubic lattice of spacing $a$ and volume 
$V=a^4 N_t\times N_s^3$ with the usual Wilson action \cite{W} 
\begin{equation}\label{action}
S=\beta\sum_{x,\, \mu<\nu}\left(1-\frac{1}{2}\Real\Tr P_{\mu\nu}(x)\right)
\end{equation}
Here $P_{\mu\nu}(x)$ is the parallel transport along the elementary loop of the lattice (plaquette) oriented along the 
$\mu\nu$-plane and positioned at the lattice site $x$. The sum is extended over all the plaquettes.

In order to generate an ensemble of configurations with the statistical distribution $e^{-S}$, with $S$ given by 
\Eqref{action}, the usual Markov chain Monte Carlo approach has been used. The updates were performed by using a mixture of 
heatbath \cite{C80, KP} and overrelaxation \cite{C87} steps in the ratio of $1$ to $5$.

The Polyakov loop $P$ is defined as the trace of the parallel transport along the compactified temporal direction, normalized 
by the number of colors. The inter-quark potential is extracted from the correlators of the Polyakov loop according to the relation
\begin{equation}\label{potential}
V(r)=-\frac{1}{N_t}\log\langle P^*(0)P(r)\rangle
\end{equation}
and the correlators of the Polyakov loops are computed by using the multilevel algorithm developed in \cite{LW01} together 
with the multihit method \cite{PPR}. 

When using the multilevel algorithm, usually the Polyakov loop correlators at all the needed distances
are computed by means of the same ensemble of configurations. This introduces strong correlations between the estimated values of 
the potential $V(r)$ calculated at different $r$'s. Because of the limited RAM available in the hardware used to perform the simulations, 
we could not adopt this strategy: different runs were used to estimate the Polyakov loop correlators at different distances.
As a consequence the error analysis is simplified, since cross-correlations are not present. Statistical errors are 
evaluated by the bootstrap method (see \emph{e.g.} \cite{NB}).  

In finite temperature gauge theories the form of the inter-quark potential predicted by the effective string picture is 
given, up to an additive normalization, by the expression (see \cite{CPR, CFPP}) 
\begin{equation}\label{V}
aV(r)=\sigma a r+\frac{D-2}{2N_t}\log\left(\frac{r}{a}\right)+\frac{c_3\,a}{r}+O(a^3r^{-3})
\end{equation}
Here $\sigma$ is the finite temperature string tension and the second term is the logarithmic correction we are 
interested in. As noted in the introduction the coefficient of the logarithmic term is independent of the microscopic 
details of the gauge theory and it is related only to the space-time dimensionality $D$ and to the temperature 
$T=1/(a N_t)$. The term $c_3a/r$ is the first non-universal correction.

The expression in \Eqref{V} has been verified with high accuracy in \cite{CFPP} for different $D=2+1$ lattice gauge theories. Since 
the $D-2$ proportionality of the logarithmic term in \Eqref{V} is a non-trivial prediction of the effective string model, 
we think it is important to numerically check \Eqref{V} also in four dimensional lattice gauge theories.

\section{The numerical results}
\label{numerics}

In order to test the effective string prediction we used two ensembles of data generated from 
lattices of various temporal extents and different coupling values.

The first ensemble, which will be denoted by $I$ in the following, is obtained by using a lattice of size $4\times 40^3$ 
and coupling $\beta_I=2.1768$, which is the deconfinement coupling on lattices with temporal extent $N_t=3$ 
(see \emph{e.g.} \cite{LTW1}) and thus corresponds to a temperature $T_I=\frac{3}{4}T_c$.

The second ensemble, denoted by $II$, is obtained by using a $6\times 40^3$ lattice at $\beta_{II}=2.2986$. This 
value of the coupling is the critical one on lattices with temporal extent $N_t=4$ (see \emph{e.g.} \cite{LTW1}), thus 
$T_{II}=\frac{2}{3}T_c$.

The collected statistics was of $10^3$ statistically independent measurement for each inter-quark distance and for each 
of the two ensembles. Numerical simulations have been performed using GRID resources provided by INFN.

\begin{table}[h]
\centering
\begin{tabular}{|l|l|l|} \hline
$r/a$ &  \hfill $aV_I(r)$ \hfill {} & \hfill $aV_{II}(r)$\hfill {} \\ \hline
4  &  1.76447(21) & 1.25000(11) \\ \hline
5  &  1.99665(30) & 1.38839(15) \\ \hline
6  &  2.21675(38) & 1.51967(19) \\ \hline
7  &  2.43000(49) & 1.64566(24) \\ \hline
8  &  2.63631(64) & 1.76808(30) \\ \hline
9  &  2.83943(61) & 1.88721(54) \\ \hline
10 &  3.0371(10)  & 2.00563(83) \\ \hline
11 &  3.2404(20)  & 2.1211(12)  \\ \hline
12 &  3.4327(40)  & 2.2357(22)  \\ \hline
13 &  3.6287(85)  & 2.3483(42)  \\ \hline
14 &  3.828(19)   & 2.4689(78)  \\ \hline
\end{tabular}
\caption{Extracted values of the potential $V(r)$ (in lattice units) for the two ensembles.}\label{V_tab}
\end{table}

The computed values of the inter-quark potential are reported in \Tabref{V_tab} and, for the reason explained in the 
previous section, all data at different values of $r/a$ are completely independent. Because of that,
in order to test \Eqref{V}, it is more convenient to fit directly the data for the potential instead of using its first derivative 
$Q(r)=\frac{\mathrm{d}V}{\mathrm{d}r}$, as is usually done to reduce the amount of the cross-correlations. 

We fitted the values of the potential according to 
\begin{equation}\label{fit_func}
aV(r)=c_0+c_1\frac{r}{a}+c_2\log\left(\frac{r}{a}\right)+\frac{c_3a}{r}
\end{equation}
and obtained the best fit parameters reported, together with the reduced $\chi^2$ of the fit, in \Tabref{fit_tab}.

Since the corrections to the potential predicted by the effective string model are given by a power series in $1/(\sigma r^2)$, where $\sigma$ is the
finite temperature string tension, in order to be consistent with the theory the fitting procedure has to be restricted to some region $r>r_{min}$. 
In previous studies (starting with \cite{CFGHP} and up to the recent \cite{CFPP}) it was shown that $r_{min}$ can be safely defined by 
$\sigma r_{min}^2=1.5$.
We verified that for both the considered ensembles the inequality $r_{min}<4a$ is
satisfied; the good agreement of the data with the fit function for $r>r_{min}$ is a self consistency check for the chosen $r_{min}$ value.

We also verified that consistent results are achieved by fitting the derivative of the potential instead 
of the potential itself, however the
estimates for the parameters obtained in this way are less precise than the ones reported in \Tabref{fit_tab}.

\begin{figure}[h]
\centering
\scalebox{0.5}{\rotatebox{0}{\includegraphics{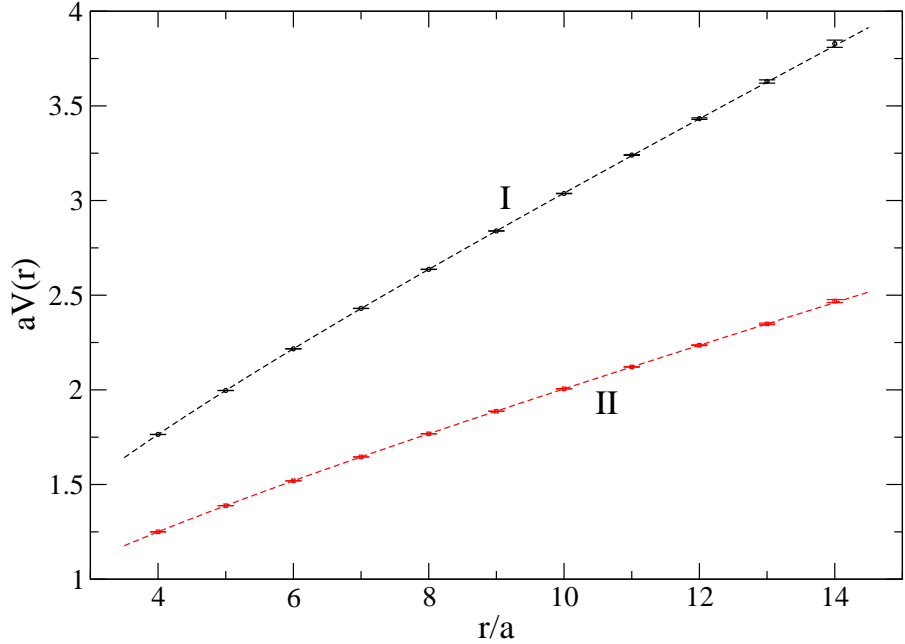}}}
\caption{Plot of the potential $V(r)$ (in lattice units) together with the fit for the two ensembles.}\label{V_fig}
\end{figure}

\begin{figure}[h]
\centering
\scalebox{0.5}{\rotatebox{0}{\includegraphics{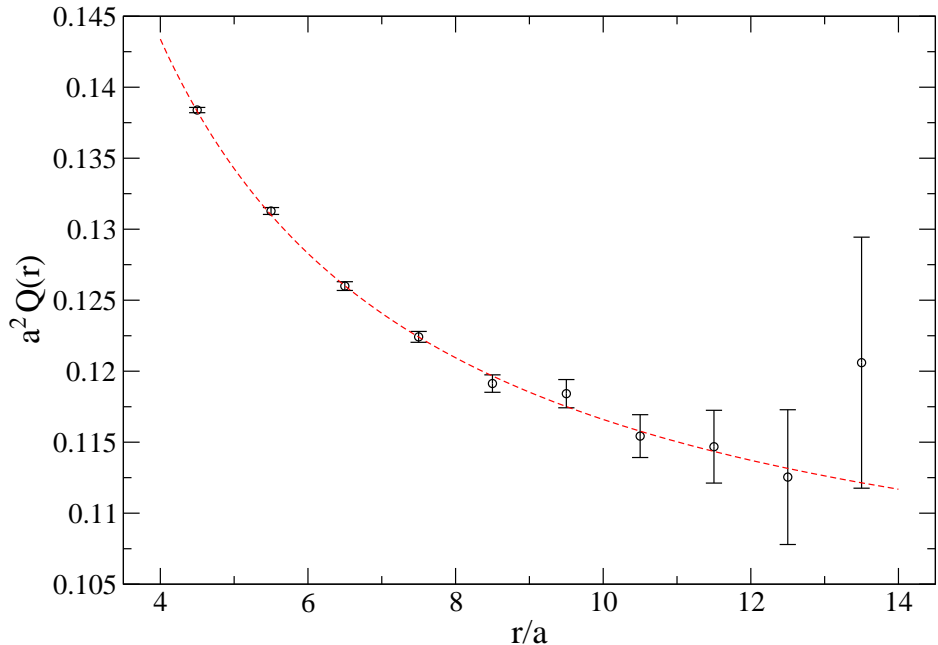}}}
\caption{Plot of $Q(r)$, the derivative of the potential, (in lattice units) together with the fit for the 
ensemble $II$.}\label{Q_fig}
\end{figure}

The data for the potential are depicted in \Figref{V_fig} together with their best fit, while in \Figref{Q_fig} the 
derivative of the potential is shown for the ensemble $II$.

\begin{table}[h]
\centering
\begin{tabular}{|c|c|c|}\hline
{}  &  $I$       &   $II$ \\ \hline
$c_0$ & 0.840(78)  &   0.631(41) \\ \hline
$c_1$ & 0.1763(35) &   0.0997(18) \\ \hline
$c_2$ & 0.200(43)  &   0.165(23)\\ \hline
$c_3$ & -0.23(13)  &   -0.037(67) \\ \hline
$\chi^2_{red}$ & 1.4 & 0.42\\ \hline
\end{tabular}
\caption{Best fit parameters for the inter-quark potential obtained by using the functional form \Eqref{fit_func}.}
\label{fit_tab}
\end{table}

The theoretical value predicted for $c_2$ by the effective string model is $c_2^{th}=\frac{D-2}{2N_t}$. For the
ensemble $I$ the theoretical expectation is thus $c_2^{th}=0.25$, which is consistent with the measured $c_2=0.200(43)$. 
For the ensemble $II$ the effective string value is $c_2^{th}=0.166$ which is in good agreement with the lattice 
estimate $c_2=0.165(23)$.

The parameter $c_1$ of the fit is the finite temperature string tension. In the Nambu-Goto effective string model this can be explicitly 
written as a function of the zero temperature string tension and it is given by (see \cite{O})
\begin{equation}\label{nambu_goto_string}
a^2\sigma=a^2\sigma_0\sqrt{1-\frac{2\pi}{3N_t^2a^2\sigma_0}}
\end{equation}
where $\sigma_0$ is the zero temperature string tension. By using the values reported in \cite{LTW1} for the string tension
at zero temperature for the couplings $\beta=2.1768$ and $\beta=2.2986$ we get from \Eqref{nambu_goto_string} the estimates
\begin{equation}\label{string_tension_ng}
\sigma_I^{(NG)}=0.1886(84) \quad \sigma_{II}^{(NG)}=0.1013(14)
\end{equation}
which are consistent with the best fit values reported in \Tabref{fit_tab}.

As a last observation we notice that the coupling used in the ensemble $I$, $\beta_I=2.1768$, is below the value $\beta=2.2$ 
which is usually reported for the so called ``bulk'' or ``roughening'' transition. Although for some lattice actions this is a real 
phase transition (see \emph{e.g.} \cite{BC}) for the $SU(2)$ lattice gauge theory with the Wilson action it is just a smooth analytical 
crossover, easily seen \emph{e.g.} in the plaquette susceptibility. This crossover signals the transition from the strongly coupled 
regime, where lattice artefacts are dominant and the strong coupling expansion is convergent, to the weakly coupled phase, where the 
continuum limit is smoothly reached (see \emph{e.g.} \cite{IPZ} and \cite{C80} for the first numerical results).

Our results concerning the ensemble $I$ show that, although it is defined only in the continuum, the 
effective string model is consistent with the behaviour of lattice gauge theories also for values of the coupling where 
strong discretization corrections are to be expected. The fact that the effective string description is compatible with the data also  
below the roughening transition is not to be seen as an internal inconsistency. Since the roughening transition is 
not, in the studied theory, a real phase transition, its location is not a well defined parameter and the departure of the effective string 
prediction from the lattice data smoothly take place in a quite large interval of coupling values.

\section{Conclusions}

The effective string model is a simple theoretical picture which encloses much of the physics of non abelian gauge theories
and give testable prediction on the behaviour of the inter-quark static potential. 

A clear signature of the effective string at zero temperature is given by the well known L\"{u}scher term. It is a recent observation that a similar role
is played, at finite temperature, by the logarithmic correction to the potential. This was up to now confirmed by lattice 
results in gauge theory only in the $2+1$-dimensional case. 

We extended these results by providing numerical 
evidence that the same is true also for the $SU(2)$ four dimensional lattice gauge theory. We also noticed that the observed behaviour of 
the potential is well described by the effective string model also at very large coupling, where large lattice artefacts are 
expected to be present. 

It can be noted that the data of the ensemble II are in better agreement with the theoretical expectations than the one of the ensemble I. This 
can in principle be related to the different temperature (higher for the ensemble I) and to the lattice artefacts (larger for the ensemble I). 
It is indeed known that the Nambu-Goto string description is not valid near the deconfinement transition (see \cite{ABT} for a 
study of the string tension in $(2+1)D$ $SU(2)$ near the deconfinement) 
and it is clear that lattice artefacts are not taken into account in the effective string picture. Further studies will be needed to gain a 
better understanding of 
this point and, more generally, of the limits of applicability of the effective string model to lattice gauge theory.

\section{Acknowledgement}
It is a pleasure to thank Michele Caselle for useful comments and encouragement.

\bibliographystyle{model1-num-names}

\end{document}